\documentclass[preprint2]{aastex}
\usepackage{graphicx}
\usepackage{amsmath}
\usepackage{amssymb}
\usepackage{cases}

\shorttitle{ICA in Exoplanet Spectroscopy}
\shortauthors{Waldmann}

\begin{document}

\title{Blind extraction of an exoplanetary spectrum through Independent Component Analysis}

\author{I. P. Waldmann, G. Tinetti}
\affil{Department of Physics \& Astronomy, University College London, Gower Street, WC1E 6BT, UK}
\email{ingo@star.ucl.ac.uk}

\author{P. Deroo}
\affil{Jet Propulsion Laboratory, California Institute of Technology, 4800 Oak Grove Drive, Pasadena, California 91109-8099, USA}

\author{M.~D.~J. Hollis, S.~N. Yurchenko, J. Tennyson}
\affil{Department of Physics \& Astronomy, University College London, Gower Street, WC1E 6BT, UK}

\begin{abstract}

Blind-source separation techniques are used to extract the transmission spectrum of the hot-Jupiter HD189733b recorded by the {\it Hubble}/NICMOS instrument. Such a `blind' analysis of the data is based on the concept of independent component analysis. The de-trending of {\it Hubble}/NICMOS data using the sole assumption that nongaussian systematic noise is statistically independent from the desired light-curve signals is presented. By not assuming any prior, nor auxiliary information but the data themselves, it is shown that spectroscopic errors only about 10 - 30$\%$ larger than parametric methods can be obtained for 11 spectral bins with bin sizes of  $\sim$ 0.09$\mu$m. This represents a reasonable trade-off between a higher degree of objectivity for the non-parametric methods and smaller standard errors for the parametric de-trending. 

Results are discussed in the light of previous analyses published in the literature. The fact that three very different analysis techniques yield comparable spectra is a strong indication of the stability of these results.

\end{abstract}


\section{Introduction}

The field of exoplanetary spectroscopy is as rapidly advancing as it is new. It has come from the first detection of spectroscopic features in an exoplanetary atmosphere \citep{charbonneau02}, to an ever more detailed characterisation of a variety of targets. 
\citep[e.g.][]{agol10, beaulieu10, beaulieu11, berta12, charbonneau05,charbonneau08, deming05, deming07,grillmair08, knutson07a, snellen08, snellen10a, snellen10b, brogi12,  bean11b,  stevenson10, swain08, swain09a,swain09b,tinetti07, tinetti10, crouzet12}. 
The aim to characterise smaller and smaller planets is equally a quest for higher and higher precision measurements, which are often limited by the systematic noise associated with the instrument with which the data are observed. 
This is particularly true for  general, non-dedicated observatories. In the past, parametric models have extensively been used by several teams to 
remove instrument systematics \citep[e.g. ][]{agol10, beaulieu10, beaulieu11, brown01b, burke10, charbonneau05,charbonneau08, deming07,desert11,gibson11,grillmair08, knutson07a, pont08, sing11, swain08,swain09b}. 
Parametric models  approximate systematic noise via the use of auxiliary information of the instrument, the so called optical state vectors (OSVs). Such OSVs often include the X and Y-positional drifts of the star or the spectrum on the detector, the focus and the detector temperature changes, as well as positional angles of the telescope on the sky. By fitting a linear combination of OSVs to the data, the parametric approach derives its systematic noise model. We refer to this as the `linear, parametric' method.
In the case of dedicated missions, such as {\it Kepler} \citep{borucki96,jenkins10}, the instrument response functions are well characterised in advance and conceived to reach the required 10$^{-4}$ to 10$^{-5}$ photometric precision.
For general purpose instruments, not calibrated to reach this required precision,  poorly sampled optical state vectors or a missing parameterisation of the instrument often become critical issues.
Even if the parameterisation is sufficient, it is often difficult to determine which combination of these OSVs may best capture the systematic effects of the instrument.

Given the intricacies of a parametric  approach, several groups have worked towards alternative methods to decorrelate the data from instrumental and stellar noise. 
The issue of time-correlated systematics in exoplanetary time series was discussed by \citet{pont06}. \citet{carter09} developed a wavelet based, non-parametric, de-trending of 1/$f$ noise contaminated lightcurves. \citet{thatte10} proposed a selective principal component filtering to reduce instrument and telluric systematics. 
More recently, \citet{gibson11b},here after G12, presented a non-parametric de-trending approach based on Gaussian Processes \citep[GP;][]{rasmussen06}.  GP, as implemented by G12, belongs to the class of non-parametric, supervised machine learning algorithms. It uses the observed data and OSVs to derive an optimal, non-linear systematic noise model for the observed data. 

In \citet{waldmann12b}, here after W12, we proposed independent component analysis \citep[ICA;][]{hyvarinen99} as an effective way to de-correlate the exoplanetary signal from the instrument and stellar noise components. ICA  belongs to the category of unsupervised machine learning algorithms (i.e. the algorithm does not need to be trained prior to use). It does not require auxiliary information such as OSVs but only the observed data themselves. This approach is also known as blind decorrelation as the assumptions on the system are minimal. As described later on, ICA assumes a linear combination of independent components (ICs) to form its systematic noise model. 

The issue of poorly constrained parameter spaces is in fact not new in astrophysics and has given rise to an increased interest in blind-source separation algorithms. 
Cosmological and extragalactic observations, in particular, are often analysed  through fully blind, non-parametric methods and ICA has successfully been used to separate the cosmic microwave background (CMB) or the signatures of distant galaxies from their galactic foregrounds \citep[e.g.][]{chapman12, stivoli06, maino02, maino07, wang10}. \citet{aumont07}, for instance,  can separate the instrumental noise from the desired astrophysical signal by using ICA. 

In W12 the ICA approach was tested by presenting two single, de-correlated light-curves of the primary transit of HD189733b and XO1b obtained with {\it Hubble}/NICMOS in its spectroscopy setting. In this paper, we apply the ICA de-trending method to the extraction of an exoplanetary spectrum; we then compare the results obtained to the previous ones published in the literature and discuss the advantages and limits of this technique for analysing exoplanetary spectra and light-curves. 

\subsection{The HD189733b {\it Hubble}/NICMOS data-set}

As explained in the introduction,  the main focus of our paper is the application and critical discussion of the ICA method to de-trend time-resolved spectroscopic data.
We chose the transit spectrum of the hot-Jupiter HD189733b recorded by {\it Hubble}/NICMOS as a testbed, the main reason being that these data have been analysed by other teams in the past and the results of the analysis have been quite debated in the literature.   

The NICMOS data were first published by S08 and then re-analysed by \citet{gibson11} using a similar linear parametric de-trending approach. 
\citet{gibson11}, failing to retrieve the transmission spectrum reported by S08,  attributed the discrepancy to a high degree of degeneracy between the systematic noise correction model used and the extracted spectrum. \citet{swain11} argued, though, this discrepancy being due to poorly derived OSVs by \citet{gibson11}.  
A more recent re-analysis of the data by G12  using non-parametric, non-linear {\it Gaussian Processes}, found a solution consistent  with that of S08, but error bars on average 2.4 times as large. 

Given these debates, it becomes critical to understand how far we can push the analysis when no prior or auxiliary information for the instrument is assumed.

\section{Data analysis}

In this section we briefly introduce the ICA technique and we apply it to the primary transit data of HD189733b, as recorded by {\it Hubble}/NICMOS. 

\subsection{Independent Component Analysis}

Let us assume we have multiple, simultaneous observations of a mixture of signals. In time-resolved spectroscopy (spectrophotometry) we obtain a light-curve signal of the exoplanetary transit per resolution element of the spectrograph. 
Here, each measured time series is denoted by $x_{k}(t)$, where $k$ is its index in the individual time series and $N$ is the total number of time series. The measured signal can be assumed to be the sum of the astrophysical light-curve signal, $s_{a}(t)$, the  instrumental and stellar noise sources, $s_{sn}(t)$ and the white noise, $s_{wn}(t)$. So:

\begin{align}
\label{intro1}
x_{k}(t) &=  a_{k,1} s_{a}(t) + s_{wn}(t) +  a_{k,2} s_{sn1}(t) +\\\nonumber
& + a_{k,3} s_{sn2}(t)  + ... + a_{k,N_{sn}} s_{sn}(t)
\end{align}

\noindent or as sum of vectors (the time-dependance has been dropped for clarity):
\begin{equation}
{x}_{k} = a_{k,1}{s}_{a} +  a_{k,2}{s}_{wn} + \sum_{l=3}^{N_{sn}} a_{k,l} {s}_{sn} 
\label{intro2}
\end{equation}

\noindent where $l$ is the estimated source signal index. For non-overcomplete sets, we have as many individual source signals as time series, i.e. $k = l$. Assuming only one source signal is astrophysical and one is Gaussian per time series, we can state the total number of source signals to be $N~=~N_{sn}~+~2$, where $N_{sn}$ is the number of systematic noise source signals. 

It is worth noting that for overcomplete sets, we have more time series available than source signals contained within, i.e. $k > l$. In these cases we can reduce the dimensionality of the data set (for example using principal component analysis) or select a sub-set of estimated source signals given some selection criteria (W12).  

We can also express eq.~\ref{intro1} in matrix form as: 

\begin{equation}
\bf{x}=\bf{A}\bf{s}
\label{intro3}
\end{equation}

\noindent where $\bf{x}$ is the column vector containing the measured time series, $x_{k}$, i.e. ${\bf x} = (x_{1}, x_{2}, \dots, x_{N})^{T}$,  $\bf{s}$ is the column vector of  independent source signals, ${\bf s} = (s_{a}, s_{wn}, s_{sn1}, s_{sn2}, \dots, s_{N})^{T}$. We may also write ${\bf s} = \bf{s_{a}} + \bf{s_{wn}} + \bf{s_{sn}}$ to clearly differentiate between the astrophysical, white noise and systematic components. 
$\bf{A}$ is the $N \times N$ dimensional `mixing matrix' comprised of the weights $a_{k,l}$. 

 The motivation of ICA is to estimate $\bf{A}$ without prior knowledge of $\bf{A}$ or $\bf{s}$ \citep{hyvarinen99}. This is achieved by making the stringent assumption that the source signals composing $\bf{s}$, $s_{k}$, are statistically independent from each other. Such an assumption is valid, as one expects the astrophysical light-curve signal to be independent in origin from systematic instrumental noise. ICA algorithms hence attempt to de-compose observed signals, $\bf{x}$, to a set of independent source components, $\bf{s}$.  To maximise said independence, several approaches have been proposed \citep[for a comprehensive summary:][]{icabook,icabook2}. Here we follow W12 and use a variant of the FastICA algorithm, which maximises statistical independence of the {\it estimated} source signals $s_{l}$ by maximising the nongaussianities of their respective probability distributions \citep{hyvarinen99,hyvarinen99b, hyvarinen00,koldovsky06, icabook, icabook2}. 
 
A linear combination of individual source signals, as in eqs.~\ref{intro1} and \ref{intro2}, is hence a valid assumption as each component is assumed to be fully independent. However, a few subtleties regarding this approach are worth mentioning. \emph{ Only the signals that are common to all time series, $x_{k}(t)$, can realistically be de-convolved.  } Taking {\it Hubble}/NICMOS as  example, systematics introduced by the grism are easily detectable, as these are `common' to all dispersed wavelengths. Similarly, detector-wide flat-fielding gradients are de-trendable. On the other hand, localised inter-pixel variations are not represented in all time series and would not be de-correlated with ICA. These properties lead to the discussion of `global' and `local' noise models later on in the text.  Whilst the effect of limb-darkening is reduced in the infra-red, it is true that wavelength varying limb-darkening coefficients can impair the direct extraction of the lightcurve feature. However, as described in the next section, the systematic noise model makes no direct use of the astrophysical component, $ \bf s_{a}$, and hence circumvents this potential limitation. 

\subsection{Application to HD189733b}
\label{sec:hd189}

The primary transit of HD189733b was observed using {\it HST}/NICMOS in the G206 grism setting and spanning five consecutive orbits. The selected grism covers the spectral range of $\sim$1.51 - 2.43 $\mu$m, see S08. The {\it HST}-pipeline calibrated data were downloaded from the MAST\footnote{http://archive.stsci.edu/} archive and the spectrum was extracted using both standard IRAF\footnote{http://iraf.noao.edu/} routines as well as a custom built routine for optimal spectral extraction. Both extractions are in good accord with each other but the custom built routine was found to yield a better signal to noise and was subsequently adopted for all further analysis. In order to minimise inter- and intra-pixel variability of the NICMOS detector, the instrument was slightly de-focused to a full-width-half-maximum (FWHM) of $\sim$5 spectral channels per resolution element. This sets a limit on the maximum resolution, $R$, achievable. 

\begin{figure}
\centering
\includegraphics[width=7.8cm, keepaspectratio=true]{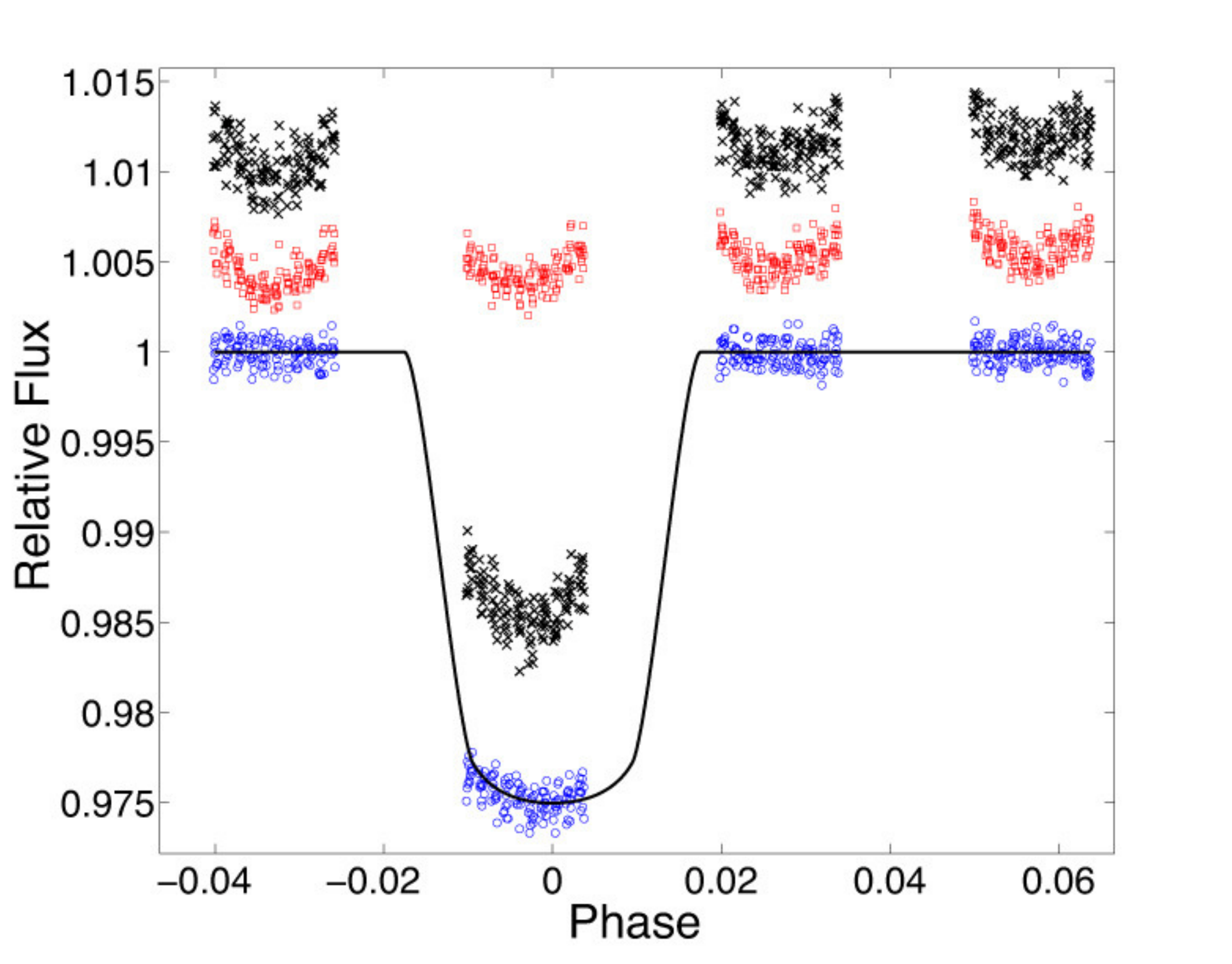}
\caption{Raw light-curve at $\sim$2.33 $\mu$m (black crosses), its respective systematic noise model (red squares), $m_{k}(t)$, composed out of the systematic components in figure~\ref{syscomp}. The de-trended final light-curve is shown underneath (blue circles) with a \citet{mandel02} fit overlaid.}
\label{modelcurve}
\end{figure}

ICA is limited by the amount of Gaussian noise present in the data \citep{hyvarinen00}. We found the minimal binning of 5 channels to be too low
 in SNR and used a binning of 8 spectral channels ($\sim$ 0.09$\mu$m). Several data pre-processing methods exist to decrease the Gaussian component of time series data (e.g. kernel smoothing, low pass filters, wavelet based approaches, etc.) but we decided to interfere as little as possible with the original data and opted for a slightly coarser binning instead. This resulted in 11 light curves across the G206 grism band. We found the first of the 5 orbits to be very noisy and negatively impacting the efficiency of the algorithm and excluded the first orbit from all further analysis. An example of the `raw' light-curves' quality, at $\sim$2.33 $\mu$m, can be found in figure~\ref{modelcurve}. 

\begin{figure}
\centering
\includegraphics[width=7.8cm]{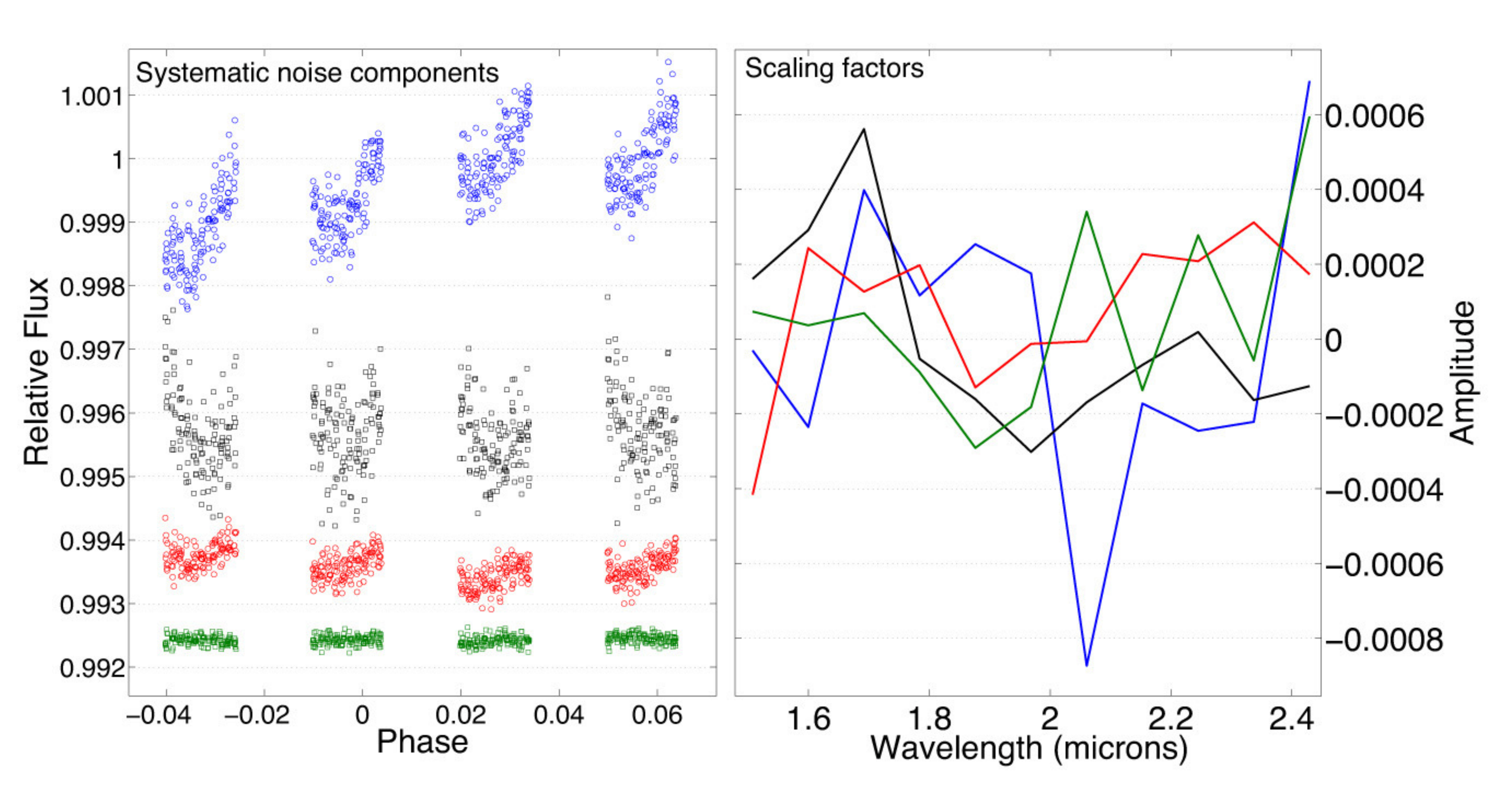}
\caption{LEFT: Four retrieved nongaussian systematic noise components in the order of importance. They were computed over the whole spectral range of the G206 grism and describe the systematic noise (instrumental and/or stellar) common to all spectral channels. RIGHT: Scaling factors, $o_{k,l}$, of the systematic noise components on the left. The colour coding is identical for both plots. We can see that the first component at 2.06~$\mu$m is sharply deviating from its own pass-band mean and the mean of all components at 2.06~$\mu$m. This can indicate that the `global' systematic noise model does not describe well the systematics in this channel and could over-correct. }
\label{syscomp}
\end{figure}

As described in W12, we used the extracted light-curves as input to the ICA algorithm to calculate the mixing matrix, $\bf{A}$, and its (pseudo)inverse, the de-mixing matrix $\bf{W} = \bf{A}^{-1}$.  Once the de-mixing matrix had been determined, the algorithm tested the estimated components for their nongaussianity and returned four main systematic noise components which do not correlate with the expected light-curve morphology. 
These components, comprising $\bf{s}_{sn}$, were extracted over the entire spectral range of the grism (referred to as `global' below) and showed a good degree of separation (figure~\ref{syscomp}, left side). 
ICA estimates the mixing matrix $\bf{A}$ up to a sign and scaling factor, meaning the source signals, $\bf{s}$, are recovered but lack an overall scaling constant. In an analogy to principal component analysis, we can think of this as recovering the eigenvectors but not the eigenvalues. Due to this ambiguity, we need to determine the scaling of the individual systematic noise components, per time series $x_{k}$, separately. This is done by fitting a systematic-noise-model (SNM), $m_{k}$, to the out-of-transit data of each time series $x_{k}$, where $m_{k}$ is the sum of all scaled systematic noise components in $\bf{s}_{sn}$, i.e. $m_{k} = \sum_{l = 1}^{N_{sn}} o_{k,l} {s}_{sn,l}$, where $o_{k,l}$ is the scaling factor for the systematic noise signal $s_{sn,l}$ for a given time series $k$.
A Nelder-Mead minimisation algorithm \citep{press07} was used to fit for $o_{k,l}$. The scaling amplitude of each component for each light-curve is given in figure \ref{syscomp} (right side). Once $m_{k}$ is determined, we subtract it from the raw data to get the corrected time series $y_{k} = x_{k} - m_{k}$, see figure~\ref{modelcurve} for an example. 

\subsection{Lightcurve fitting and Error-bars}

Having obtained the de-trended time series, $y_{k}$, we proceed to model-fit these using the analytical model by \citet{mandel02}, from here MA02, with orbital and limb-darkening parameters taken from S08. The S08 quadratic limb-darkening parameter are interpolated to the coarser wavelength grid of this analysis. The transit depth, $\delta_{k}$, is left as only free parameter. The transit depths of all 11 light-curves constitute the exoplanetary spectrum. The model-fitting was performed using a Markov Chain Monte Carlo (MCMC) algorithm and cross checked using two variants of a Bootstrap Monte Carlo analysis (see appendix~\ref{appendix:boot}).

\subsubsection{Markov Chain Monte Carlo}
\label{sec:mcmc}

MCMC \citep{press07} has become the standard fitting routine for exoplanetary time series and radial velocity data \citep[e.g.][]{ford06,burke07,bakos07,knutson07a,cameron07,charbonneau09, bean11, kipping11,gregory11,crouzet12,knutson12}. In this analysis we use an adaptive version of the standard Metropolis-Hastings algorithm \citep{haario01, haario06, metropolis53, hastings70, press07}. The only free parameter, $\delta_{k}$,  was set to have a uniform prior ranging from $R_{p}/R_{\ast}$ = 0~-~1. As we only fit for the transit depth we are not concerned by inter-parameter correlations in this analysis. 

We perform an initial MA02 model fit, $m_{MA02,k}(t)$, using a Nelder-Mead minimisation algorithm \citep{press07} and calculate the model subtracted residual, $r_{k}(t) = x_{k}(t) - m_{MA02,k}(t) $. We take the fitted transit depth as starting value of the MCMC chain and calculate the variance of the normal sampling distribution as

\begin{equation}
\sigma_{k,sample}^{2} = var(r_{k}(t)) + 2 \sum_{\tau =1} cov[r_{k}(t),r_{k}(t+\tau)]
\label{likelihood}
\end{equation}

\noindent where $var$ is the variance of the residual and $cov$ the auto-covariance for a given lag $\tau$. This accounts for remaining autocorrelated noise in the time series data. The MCMC algorithm was consequently run for $2\times 10^{4}$ iterations to guarantee a good coverage of the posterior distribution, of which examples are shown in figures~\ref{errplot1}a \& \ref{errplot2}a. Here the error bars are the standard deviation of the posterior distribution. 

In addition to the MCMC algorithm described here, we also estimated the standard error using two variants of a Bootstrap Monte-Carlo algorithm, see appendix~\ref{appendix:boot}. We find the retrieved transit-depths and errors to be in good agreement with the MCMC method.  

\subsubsection{Source signal separation error}

The two core algorithms used in this analysis EFICA \citep{koldovsky06} and WASOBI \citep{yeredor00, tichavsky06b} can be shown to be asymptotically efficient, i.e. reaches the Cramer-Rao Lower Bound (CRLB) in an ideal case where the nonlinearity $G(.)$ equals the signal's score function. In other words, the algorithms employed here can be shown to converge to the correct solution given the original source signals and in the limit of $N_{iter} \rightarrow \infty$ iterations. 

In reality the number of iterations is finite and imperfect convergence results in traces of other sources to remain in the individual signals comprising $\bf{s}$.  We can hence state that the estimated de-mixing matrix, $\bf{W}$ is only approximately equal to the inverse of the original mixing matrix, $\bf{A}$, i.e. 

\begin{equation}
\bf{W} \simeq \bf{A}^{-1}
\label{demix2}
\end{equation}

\noindent This requires us to calculate the signal separation error (SSE) of the analysis. A measure of this error is the deviation of $\bf{WA}$ from the unity matrix by inspecting the variance of its elements \citep{koldovsky06,icabook}. 

To assert a good degree of separation, we can define $\bf{G}$ as the gain matrix. For a perfectly estimated de-mixing matrix, $\bf{W}$, the gain matrix is equal to its identity matrix  

\begin{equation}
\textbf{G} = \textbf{W}\textbf{A} = \textbf{I} 
\label{gainmatrix}
\end{equation}

In signal processing, the performance of blind-source separation algorithms is usually measured by the signal over inference ratio, SIR\footnote{standard literature also refers to its inverse, the interference over signal ratio, ISR}. The SIR is the standard measure in signal processing of how well a given signal has been transmitted or de-convolved from a mixture of signals. Given the inference to be a noise source, we can equate this to the more commonly used signal-to-noise ratio (SNR). We can now calculate the separation error of the estimated source signal, $s_{l}$, in relation to the original source signal, $s_{k}$, using

\begin{equation}
\sigma_{l} = \frac{{\text E}[\sum^{N}_{l=1,l \neq k} \textbf{G}^{2}_{kl}]} {{\text E}[\textbf{G}^{2}_{kk}]}, ~ k ,l = 1,2,...,N.
\label{isrk}
\end{equation}

However, the original mixing matrix, \textbf{A}, and the original source signals, $s_{k}$, are not generally known for real data sets and equation \ref{isrk} is only useful in the case of simulations. \citet{tichavsky06,koldovsky06} and \citet{tichavsky08} have shown that despite the original mixing matrix being unknown,  an asymptotic estimate of the SIR can be made. A derivation of this process is beyond the scope of this paper and we refer the interested reader to the relevant literature. 

We now have the retrieval error on the individual source components, $\sigma_{l}$. In order to obtain the overall error on the systematic noise model, $m_{k}$, calculated in section~\ref{sec:hd189}, we compute the weighted sum of the individual source separation errors with the previously retrieved source component weighting factors, $o_{k,l}$

\begin{equation}
\sigma_{k,SSE} = \left(\sum_{l = 1}^{Nsn}  o_{k,l}^{2}\sigma^{2}_{l}\right)^{1/N_{sn}}
\label{snmodelerr}
\end{equation}

\subsubsection{SNM fitting error}

In addition to the above determined errors, we also include an ICA fitting error which accounts for possible over-corrections of the global SNM to individual, poorer constraint light-curves. This term becomes non-zero when the scaling of a systematic noise component, $o_{k,l}$ (figure~ \ref{syscomp}, right hand side), shows a 3$\sigma$ significant deviation from the mean scaling of all other light-curves, $\bar{o}_{k}$. In other words, we expect the scaling of an individual systematic noise component, $\bf {s}_{sn}$, to be a slowly varying function over wavelength for `globally' estimated systematic-noise components. If individual light-curves show a significantly larger positive or negative scaling than expected, for an individual lightcurve, we can assume the nongausian noise of the affected light-curve not to be properly accounted for by this global model. In larger data-sets it is easier to exclude the affected light-curve from any further analysis, whilst in small data-sets we take the amplitude of the scaling from its mean scaling as the error, i.e. 

\begin{equation}
\sigma_{k,SNM} = |(o_{k,l} - \bar{o}_{k})|
\label{icafiterr}
\end{equation}

In this analysis we find the ICA fitting error to be zero for all wavelengths but the 2.06$\mu$m spectral point. 

\subsubsection{Final Error Bar}

In summary the final error bar per time series, $k$, consists of: 

\begin{enumerate}
\item \textbf{Standard error: } Estimating the variance in retrieved transit depth when model fitting of the de-trended lightcurves.
\item \textbf{Signal separation error: } Estimating the ICA component separation error.
\item \textbf{Systematic noise model error: } Estimating errors due to noise model over- or under-fitting for individual time series. 
\end{enumerate}

We now define the final error to be the sum of squares of the above mentioned error sources:

\begin{equation}
\sigma_{k,TOTAL} = \sqrt{\sigma_{k,MCMC}^{2}+ \sigma_{k,SSE}^{2} + \sigma_{k,SNM}^{2}}
\end{equation}

\section{Results}
\label{sec:results}

\begin{figure*}
\centering
\includegraphics[width=15cm]{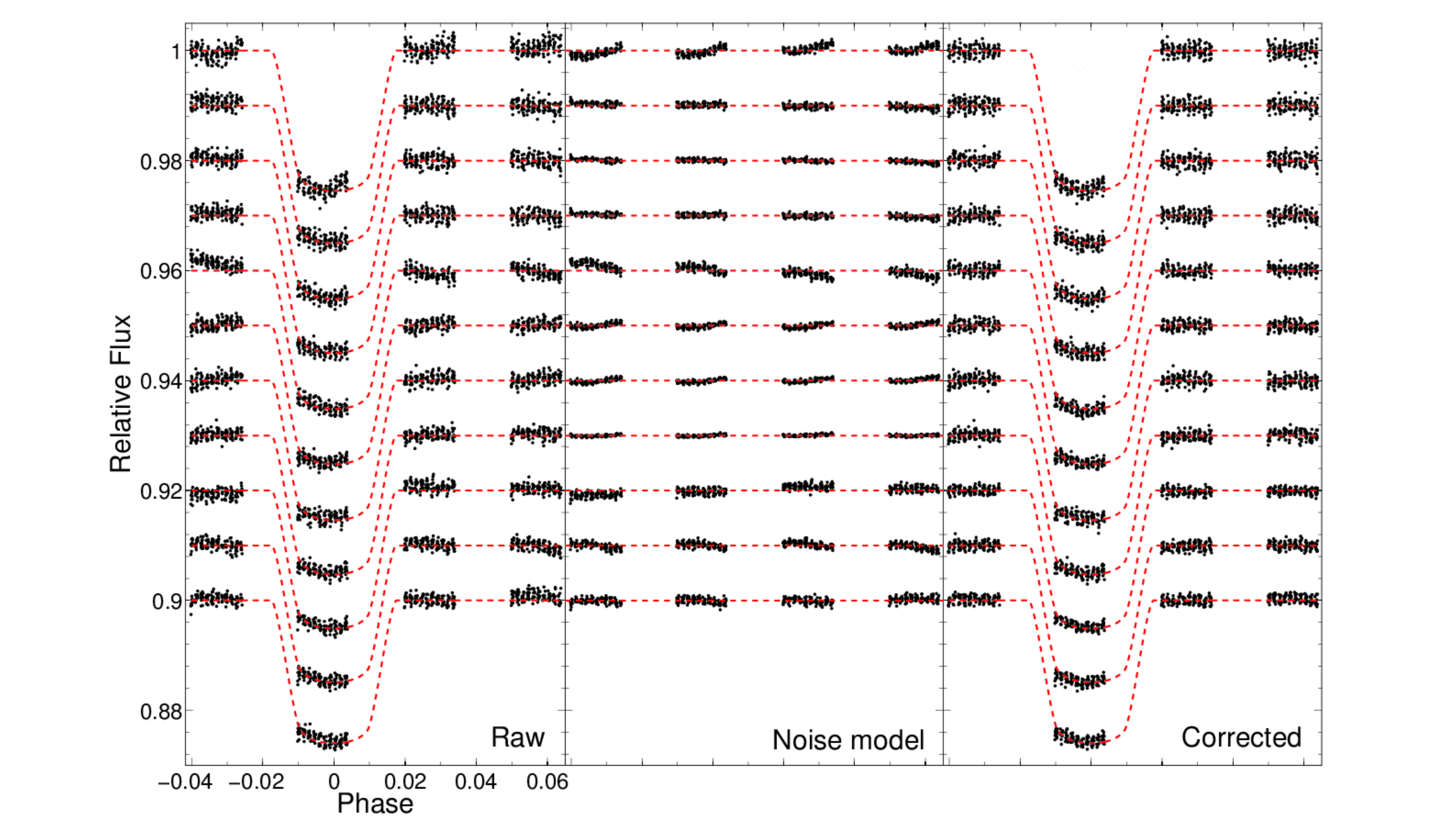}
\caption{Left: Raw light-curves from 1.51$\mu$m (bottom) to 2.43$\mu$m (top) with fitted \citet{mandel02} model overlaid. The lightcurves were off-set for clarity. Middle: Systematic noise model for each raw lightcurve in the left panel. Right: Final de-trended light-curves with fitted \citet{mandel02} model overlaid.   }
\label{lcs}
\end{figure*}

\begin{figure*}
\centering
\includegraphics[width=13cm, keepaspectratio=true]{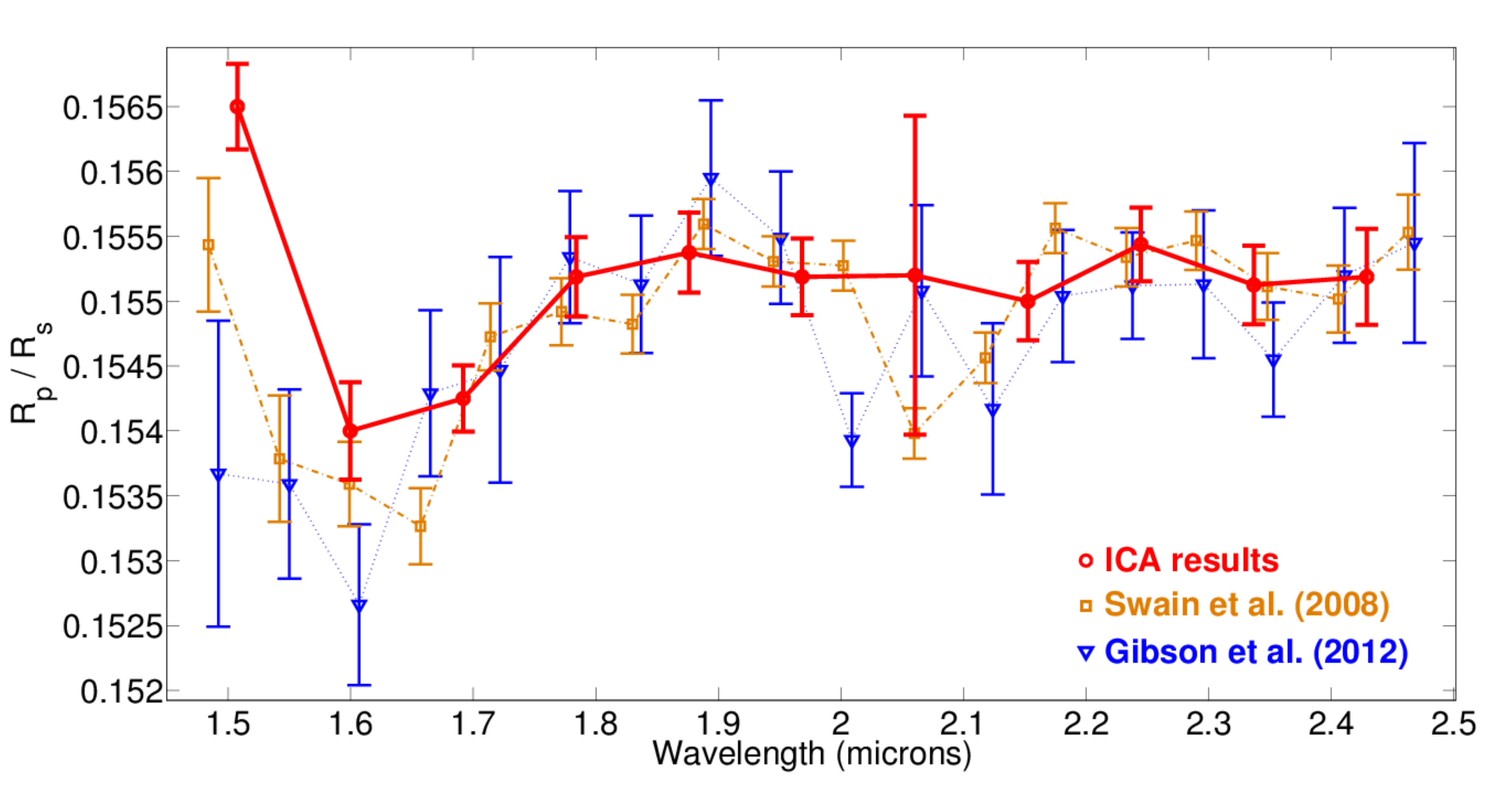}
\caption{Final spectrum (red circles) obtained with the ICA algorithm described here and in \citet{waldmann12b}, overlaid on the ersults of  \citet{swain08} (squares)  and G12 (triangles).}
\label{spectrum}
\end{figure*}

Figure~\ref{modelcurve} shows the raw light-curve at $\sim$2.33$~\mu$m in black crosses with its corrected counterpart (blue circles) offset below. Here, much of the autoregressive noise in the original data could be captured by the noise model (red squares) and removed from the final result. All raw and  corrected light-curves are shown in figure~\ref{lcs} with their respective systematic noise models. The resulting  spectrum is presented in figure~\ref{spectrum} and table~\ref{resultstable}. We find the retrieved spectrum to  be in good agreement with the `parametric' analysis S08 and G12 in terms of spectral shape which  show-cases the robustness of this methodology and the stability of the result as a whole. 

The underlying noise of the spectral point at $\sim$2.06~$\mu$m was flagged by the algorithm to be discrepant with the global systematic noise model. This can also be seen by the higher systematic noise remaining in the corrected lightcurve (5$^{th}$ from the top in figure~\ref{lcs}). Here the first systematic noise component is indicative of an overcorrection which is reflected in the error-bar as described in the previous section. Overall, the error-bars reported here are $\sim$ 10 - 30$\%$ larger than those reported by S08 but $\sim$ 10 - 50$\%$ smaller than those reported by G12. It should be noted that this analysis uses a slightly coarser bin size yielding 11 data points whilst the S08 and G12 analyses yield 18 spectral points.

\begin{figure}[t]
\centering
\includegraphics[width=7.8cm, keepaspectratio=true]{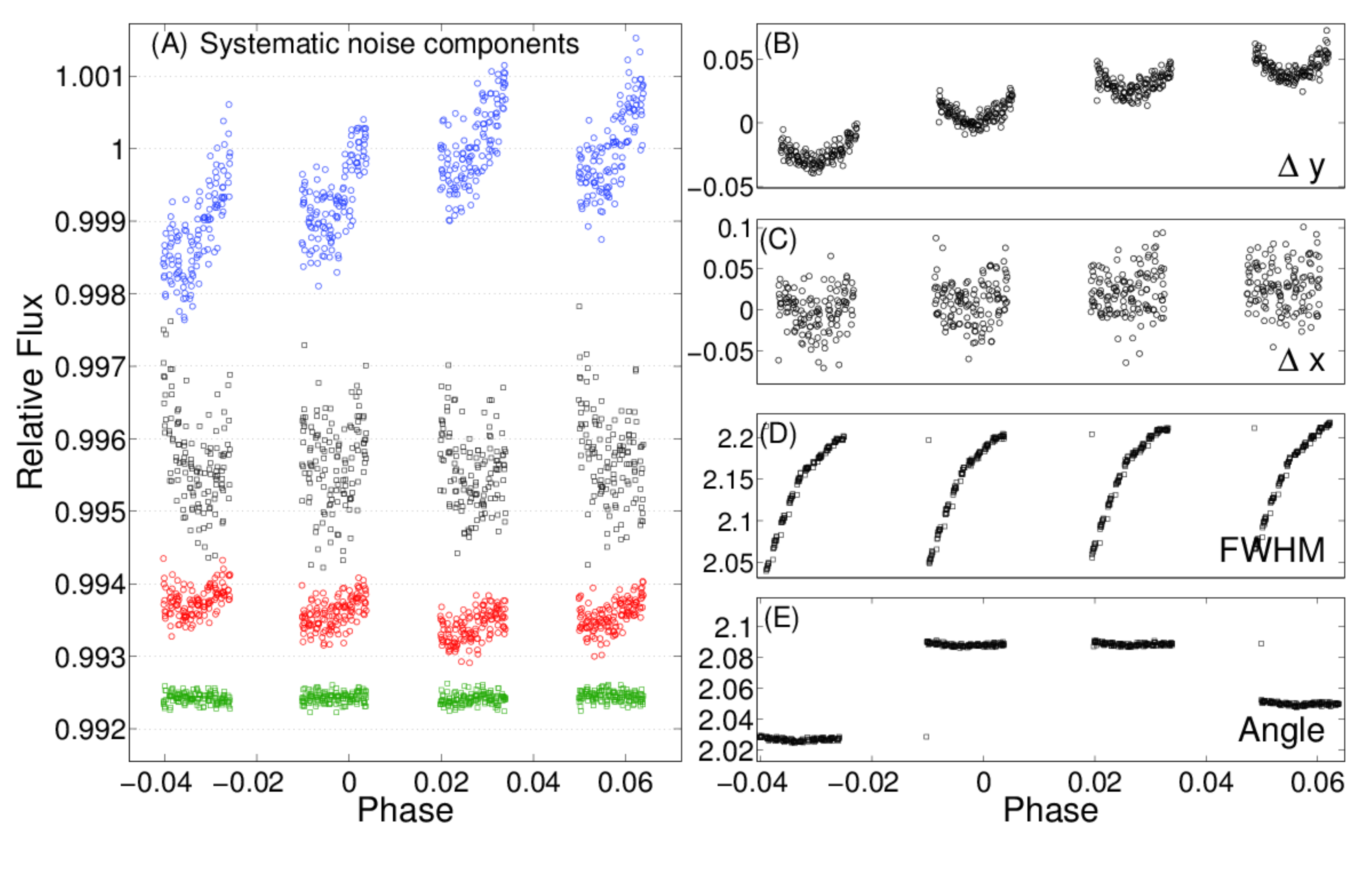}
\caption{Comparision between ICA derived nongaussian components on the left with conventionally derived OSVs as found by \citet{swain08}.  }
\label{syscomp2}
\end{figure}

\begin{deluxetable}{c c c}
%\tablecolumns{3}
\tablewidth{0pt}
\tablecaption{NICMOS transmission spectrum of HD189733b for a `global' systematic noise model correction and plotted in figure~\ref{spectrum}. The columns are wavelength ($\lambda$), planet-star-ratio ($R_{p}/R_{s}$)  and the respective error-bar ($\sigma (R_{p}/R_{s})$).\label{resultstable}}
\tablehead{\colhead{$\lambda$ ($\mu$m)} & \colhead{$R_{p}/R_{s}$} & \colhead{$\sigma (R_{p}/R_{s}) \times 10^{-4}$}}
%\tablehead{This publication & Gibson et al. (2012) & Swain et al. (2008)}
\startdata
   2.429  & 0.155187 & 3.69383 \\
   2.336  & 0.155125  & 3.03754\\
   2.244   & 0.155437   &2.83050\\
   2.152  & 0.155000  & 3.01554\\
   2.060  & 0.155200 &  12.29019\\
   1.968  & 0.155187 &  2.94831\\
   1.876  & 0.155375 &  3.09120\\
   1.784  & 0.155187 &  3.05973\\
   1.691  & 0.154250 &  2.54556\\
   1.599  & 0.154000 &  3.75548\\
   1.507 & 0.156500 &  3.28963\\
\enddata
\end{deluxetable}

\section{Discussion}
\label{discussion}

In this analysis we have computed the global SNM and found a good agreement with previously published results. Figure~\ref{spectrum} shows the comparison between the spectrum derived in this analysis, compared to the linear, fully parametric approach by S08 and the non-linear, non-parametric,  Gaussian Processes, approach by G12. Both non-parametric approaches yield slightly higher error bars than the parametric approach. We find the error on the ICA derived spectrum to be $\sim$10 - 30$\%$ bigger than those reported by S08 using a coarser bin size of $\sim 0.09 \mu$m. These differences in error bars between the `blind' and `informed' (meaning the use of auxiliary information of the instrument) approaches are not surprising. By not assuming any knowledge of the data or instrument, we are actively neglecting auxiliary information helpful to the de-trending of the data set.

With the {\it linear, non-parametric} blind ICA method presented here, the uncertainties grow by up to 30 $\sim$ 40$\%$ compared to the {\it linear, parametric} analysis by S08 (accounting for the larger bin sizes in this analysis) and a further $\sim$ 70$\%$ when further relaxing the linear assumptions as for the {\it non-linear, non-parametric} Gaussian Processes (G12).  In other words, we are trading smaller error-bars for a higher degree of objectivity.

In figure~\ref{syscomp2} we show a comparison of the ICA derived nongaussian systematic noise components and parametric OSVs. S08 identified the X and Y-positional drifts on the detector to constitute the most important OSVs in their de-correlation whilst other OSVs are less significant. Similar independent components are also present in this analysis, whilst the other independent components differ more significantly from the parametric approach.  
The agreement between the results of this analysis and those available in the literature, showcase the stability of this exoplanetary spectrum. 

We find the global SNM approach to be most sensitive to slowly varying systematic trends across the data-set while local nongaussian deviations tend not to be captured. It is therefore possible to generate, additionally,  a `local' SNM for a sub-set of spectral bins or before binning on the individual `raw' spectral channels, should the SNR permit it. It is important to remember that $k \geq N_{sn} + 1$ as the input to the algorithm. In other words, at least as many observed time series, $x_{k}$, are required as input to the algorithm than total number of nongaussian components in the data. In this case, the minimum `local' SNM would include 5 spectral bins (4 systematic noise and 1 astrophysical component). With 11 spectral bins in total, the NICMOS data-set is too small for this approach. However, for larger sets, this two stage `global' + `local' detrending becomes a viable solution. 
Furthermore, multiple observations of a transit/eclipse event with the same instrumental setup can be helpful to further de-correlate the observed data. We expect in fact the astrophysical signal being stable throughout consecutive transits, while the instrumental noise being largely uncorrelated between observations \citep{waldmann11}.

\section{Conclusion}

Here we present a reanalysis of a HD189733b primary eclipse spectrum from $\sim$1.51 - 2.43~$\mu$m obtained with {\it Hubble}/NICMOS. This analysis differs from previous publications in that it uses  blind machine-learning to de-trend data with no prior or auxiliary information about the data or instrument.  Such blind-source de-convolution algorithms can be used alone or in conjunction with other de-correlation techniques, making them very powerful tools in the analysis of exoplanetary data. This is especially true for instruments that lack an {\it a priori} calibration plan at the level of 10$^{-4}$ photometric precision, as needed for this field. 

We compare our results with previously published analyses of the same data set: another non-parametric, non-linear approach (G12) and the classical linear parametric method \citep{swain08}. We find that the error-bars of this analysis are $\sim10$ - $30\%$ larger than those reported by \citet{swain08}. We attribute this difference to the higher amount of auxiliary information injected in the parametric approach. Ultimately, it is a trade-off between a higher degree of objectivity for the non-parametric methods and smaller errors for the parametric de-trending. Additional observations would have allowed much smaller error bars and a more robust  determination of the signal at $\sim$2.06~$\mu$m. 

The fact that three very different analysis techniques yield comparable spectra is a strong indication of the stability of these results. 
The error bars estimated in this paper through ICA are smaller than the ones of \citet{gibson11b} through Gaussian Processes, suggesting more investigation is needed to identify the most effective techniques to de-trend exoplanet atmosphere data and, most importantly, understand their limitations.

\acknowledgments

The authors thank Mark Swain and Filipe Abdalla for helpful discussions. This work is supported by ERC Advanced Investigator Project 267219, STFC, NERC, UKSA, UCL and the Royal Society. All of the data presented in this paper were obtained from the Multimission Archive at the Space Telescope Science Institute (MAST). STScI is operated by the Association of Universities for Research in Astronomy, Inc., under NASA contract NAS5-26555. Support for MAST for non-{\it HST} data is provided by the NASA Office of Space Science via grant NNX09AF08G and by other grants and contracts.

\appendix

\section{Bootstrap Monte Carlo}
\label{appendix:boot}

\begin{figure}
\centering
\includegraphics[width=7.8cm, keepaspectratio=true]{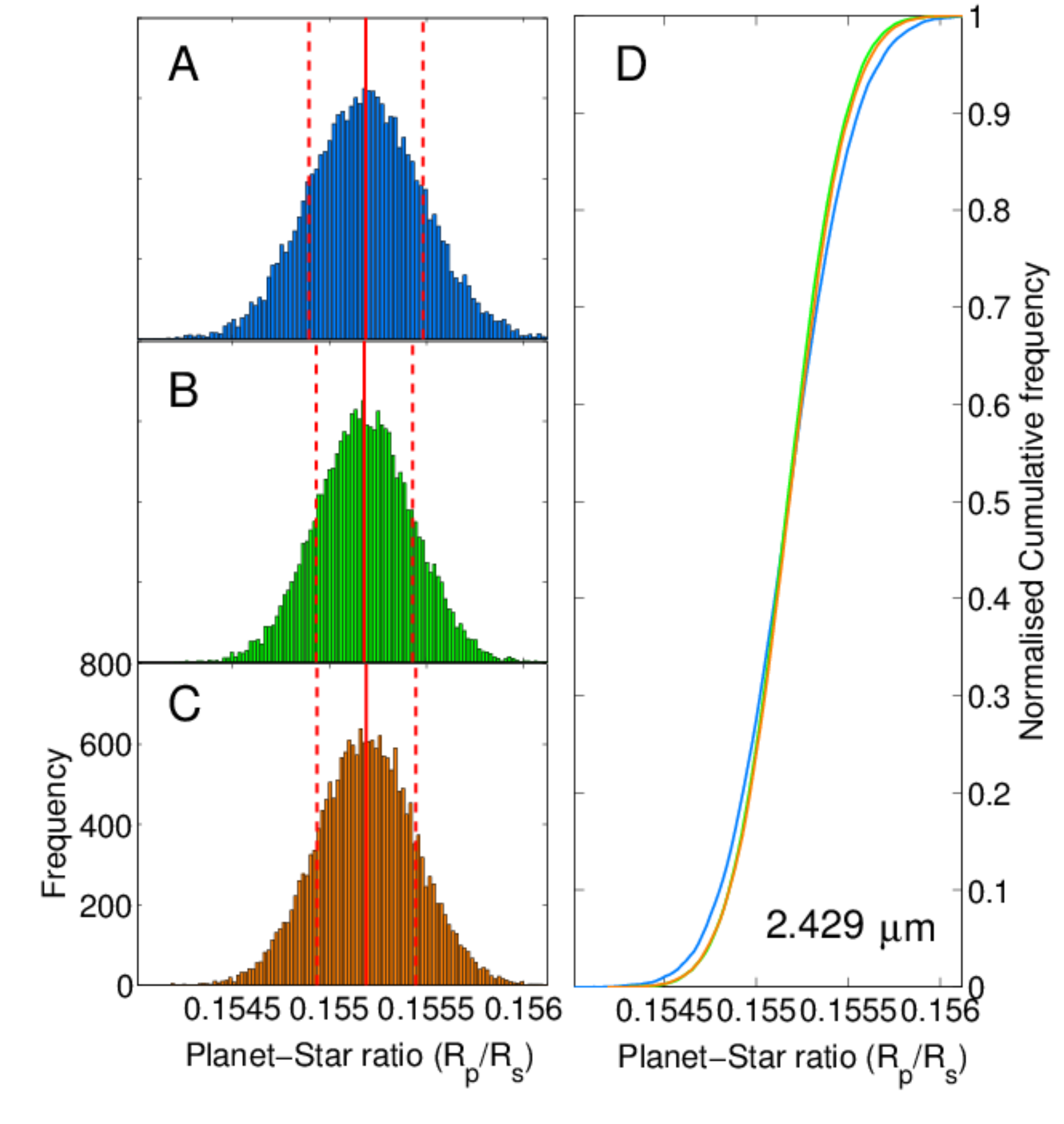}
\caption{Posterior distribution of fitted transit depth at 2.4291$\mu$m using: A) MCMC algorithm; B) Bootstrap Monte-Carlo, method 1; C) Bootstrap Monte-Carlo, method2. The sample median is marked with red, continuous line, the 1$\sigma$ error bars are marked with red-discontinuous lines. D) cumulative distribution functions of sampling distributions in plots A, B \& C.  }
\label{errplot1}
\end{figure}

\begin{figure}
\centering
\includegraphics[width=7.8cm, keepaspectratio=true]{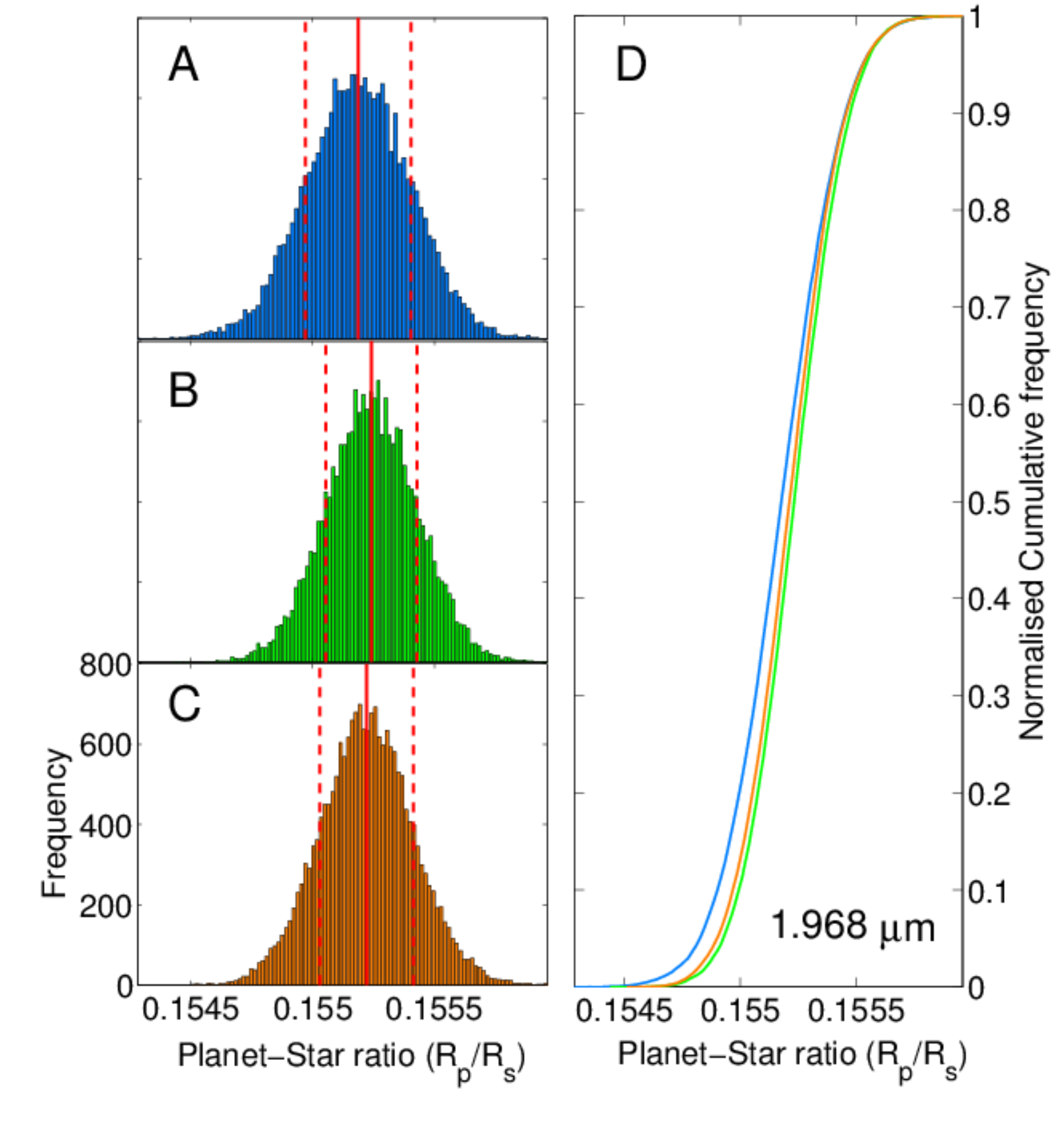}
\caption{Posterior distribution of fitted transit depth at 1.9683$\mu$m using: A) MCMC algorithm; B) Bootstrap Monte-Carlo, method 1; C) Bootstrap Monte-Carlo, method2. The sample median is marked with red, continuous line, the 1$\sigma$ error bars are marked with red-discontinuous lines. D) cumulative distribution functions of sampling distributions in plots A, B \& C.  }
\label{errplot2}
\end{figure}

Following from section~\ref{sec:mcmc}, an alternative way of estimate a parameter's sampling distribution is to use so called `bootstrap' or `jacknife' Monte-Carlo methods \citep{marcy05,press07, baluev09,baluev12}.  These methods estimate a parameter's distribution by replacing parts of the original data with a randomly permeated version of the data to observe the effect on a consequent model fit. For a given iteration of the algorithm we model-fit the time series and randomly scramble the model-subtracted residual. We then add the scrambled residual back to the original model-fit and repeat the process. Please see Appendix~\ref{bootmethod1} for details of the iteration scheme. As for the MCMC algorithm, we ran the bootstrap Monte-Carlo for $2 \times 10^{4}$ iterations and examples of the parameter distributions are found in figures~\ref{errplot1}b \& \ref{errplot2}b. It is worth noting that scrambling process destroys any autocorrelation in the data as well as homogenises nongaussian systematics in the data. 
%This said, the effect of systematic (nongaussian) noise is reflected in the increased standard deviation of the scrambled residual and hence propagates indirectly into the error-bars. 

An approach to directly measure the effects of autocorrelation in the data, the `prayer-beed' algorithm has been suggested \citep{jenkins02,southworth08}. Whereas bootstrap methods randomly replace parts of the original data, the `prayer-beed' algorithm shifts the model subtracted residual along the time-axis for every iteration. Given the small number of the data points available in this analysis and the duration of the systematics being on the same or similar time scales to the transit signal, we refrain from using this method. Instead, we re-run the above bootstrap Monte-Carlo method with the modification of only replacing a randomly sized fraction, ranging from 40-100$\%$, of the data. This modification preserves parts of the autocorrelation whilst the high iteration number of $2 \times 10^{4}$ ensures a sufficient sampling. The iteration scheme is described in Appendix~\ref{bootmethod2} and figures~\ref{errplot1}c \& \ref{errplot2}c are examples of the transit-depth sampling distributions. 

\subsection{Method 1:}
\label{bootmethod1}
\begin{enumerate}
\item Set $y_{c} = y_{k}$, where $c$ is the bootstrap iteration index. 
\item Using a Nelder-Mead minimisation and the MA02 model, evaluate and record the best fit transit-depth, $\delta_{c}$.
\item Compute the model subtracted residual, $r_{c} = y_{c} - m_{MA02}(\delta_{c})$, where $m_{MA02}(\delta_{c})$ is the MA02 model with the fitted transit depth.
\item Randomly scramble the residual to obtain $r_{c, permutated}$.
\item And add the scrambled residual back on the model to obtain the new time series $y_{c+1} = m_{MA02}(\delta_{c}) + r_{c, permutated}$
\item Steps 2 - 5 are repeated $N_{boot}$ times. 
\end{enumerate}

\subsection{Method 2:}
\label{bootmethod2}
In the second method, we follow the procedural sequence of Method~1 but only 

\begin{enumerate}
\item Set $y_{c} = y_{k}$, where $c$ is the bootstrap iteration index. 
\item Using a Nelder-Mead minimisation and the MA02 model, we evaluate and record the best fit transit-depth, $\delta_{c}$.
\item Compute the model subtracted residual, $r_{c} = y_{c} - m_{MA02}(\delta_{c})$, where $m_{MA02}(\delta_{c})$ is the MA02 model with the fitted transit depth.
\item Randomly scramble the residual to obtain $r_{c, permutated}$.
\item Randomly replace a fraction of the original residual, $r_{c}$, with the permutated residual, $r_{c,permutated}$. This fraction is chosen at random but held to be within 40 - 100$\%$ of the original data. We call this semi-permutated residual $r_{c, semi}$.
\item Add the above residual back on the model to obtain the new time series $y_{c+1} = m_{MA02}(\delta_{c}) + r_{c, semi}$
\item Steps 2 - 6 are repeated $N_{boot}$ times. 
\end{enumerate}

\end{document}